\documentclass[preprint,superscriptaddress]{revtex4}

\usepackage{amssymb}
\usepackage{amsmath}
\usepackage{natbib}
\usepackage{graphicx}
\usepackage[version=3]{mhchem}
\renewcommand{\figurename}{Fig.}

\begin{document}

\preprint{}

\title{High-Energy Quasiparticle Injection into Mesoscopic Superconductors}

\author{Loren D. Alegria}

\affiliation{Department of Physics, Harvard University, Cambridge, Massachusetts 02138}

\author{Charlotte G. L. B\o ttcher}

\affiliation{Department of Physics, Harvard University, Cambridge, Massachusetts 02138}

\author{Andrew K. Saydjari}

\affiliation{Department of Physics, Harvard University, Cambridge, Massachusetts 02138}

\author{Andrew T. Pierce}

\affiliation{Department of Physics, Harvard University, Cambridge, Massachusetts 02138}

\author{Seung Hwan Lee}

\affiliation{Department of Physics, Harvard University, Cambridge, Massachusetts 02138}

\author{Shannon P. Harvey}

\affiliation{Department of Physics, Stanford University, Stanford, California 94025}

\author{Uri Vool}

\affiliation{Department of Physics, Harvard University, Cambridge, Massachusetts 02138}

\author{Amir Yacoby}

\affiliation{Department of Physics, Harvard University, Cambridge, Massachusetts 02138}

\date{\today}

\maketitle

\textbf{
At nonzero temperatures, superconductors contain excitations known as Bogoliubov quasiparticles. The mesoscopic dynamics of quasiparticles inform the design of quantum information processors, among other devices. Knowledge of these dynamics stems from experiments in which quasiparticles are injected in a controlled fashion, typically at energies comparable to the pairing energy \cite{Levine1968,Smith1975,Ullom2000,Barends2008,Patel2017}. Here we perform tunnel spectroscopy of a mesoscopic superconductor under high electric field. We observe quasiparticle injection due to field-emitted electrons with $\mathbf{10^6}$ times the pairing energy, an unexplored regime of quasiparticle dynamics. Upon application of a gate voltage, the quasiparticle injection decreases the critical current and, at sufficiently high electric field, the field-emission current ($<$ 0.1 nA) switches the mesoscopic superconductor into the normal state, consistent with earlier results \cite{DeSimoni2018}. We expect that high-energy injection will be useful for developing quasiparticle-tolerant quantum information processors, will allow rapid control of resonator quality factors, and will enable the design of electric-field-controlled superconducting devices with new functionality.}

    Traditional superconducting information processing devices are actuated by substantial currents ($\gtrapprox\mu$A), including cryotrons\cite{Buck1956}, transmons\cite{Krinner2019}, single flux quantum processors \cite{Likharev1991}, and others\cite{McCaughan2014,Gray1978}.  The challenges of scaling transmon-style qubits to useful numbers of bits motivate recent work to actuate superconductivity by electric field effects, such as depletion of a proximally superconducting semiconductor\cite{Boettcher2018}.  But, for resonator devices, almost any added semiconductor layer deteriorates the quality factor of the device\cite{Casparis2018}.   An all-metal approach seemed improbable\cite{Glover1960} until recently, when a field effect on the critical current of metallic superconducting nanowires was observed \cite{DeSimoni2018}, but not satisfactorily explained.

	To investigate the origin of the effect we colocate a gate electrode and a local tunnel probe, as shown in Figure 1a.  Compared to the previous critical current measurements, tunnel spectroscopy provides a static measurement that can access the details of the electronic states, and, in the case of a superconductor, specifically measures the quasiparticle (QP) population\cite{Wolf2012}.

	Quasiparticle dynamics have long been a subject of fundamental inquiry.  Early studies observed how QP recombine, emitting phonons which can excite further QP until the phonons escape\cite{Levine1968,Smith1975}.  More recently, the non-galvanic propagation of QP via phonons has been spatially quantified using resonators as detectors\cite{Patel2017}.  Accelerating progress in quantum information has made QP dynamics a critical topic, as their often-uncertain populations may limit the performance of quantum processors\cite{Serniak2018}.  In such circuits, the significant presence of high-energy excitations has prompted recent use of radioisotopes as experimental sources of QP\cite{Vepsalainen2020}.  In addition, the local evolution of QP populations underpins the operation of superconducting particle detectors\cite{Li2003}, and recent progress in microfabricated particle accelerators may ultimately present opportunities for high-energy superconducting microelectronics\cite{Sapra2020}.  A conveniently fabricated, high-energy QP source is therefore of practical and fundamental interest.

\begin{figure}[h!]
\includegraphics[width=4 in]{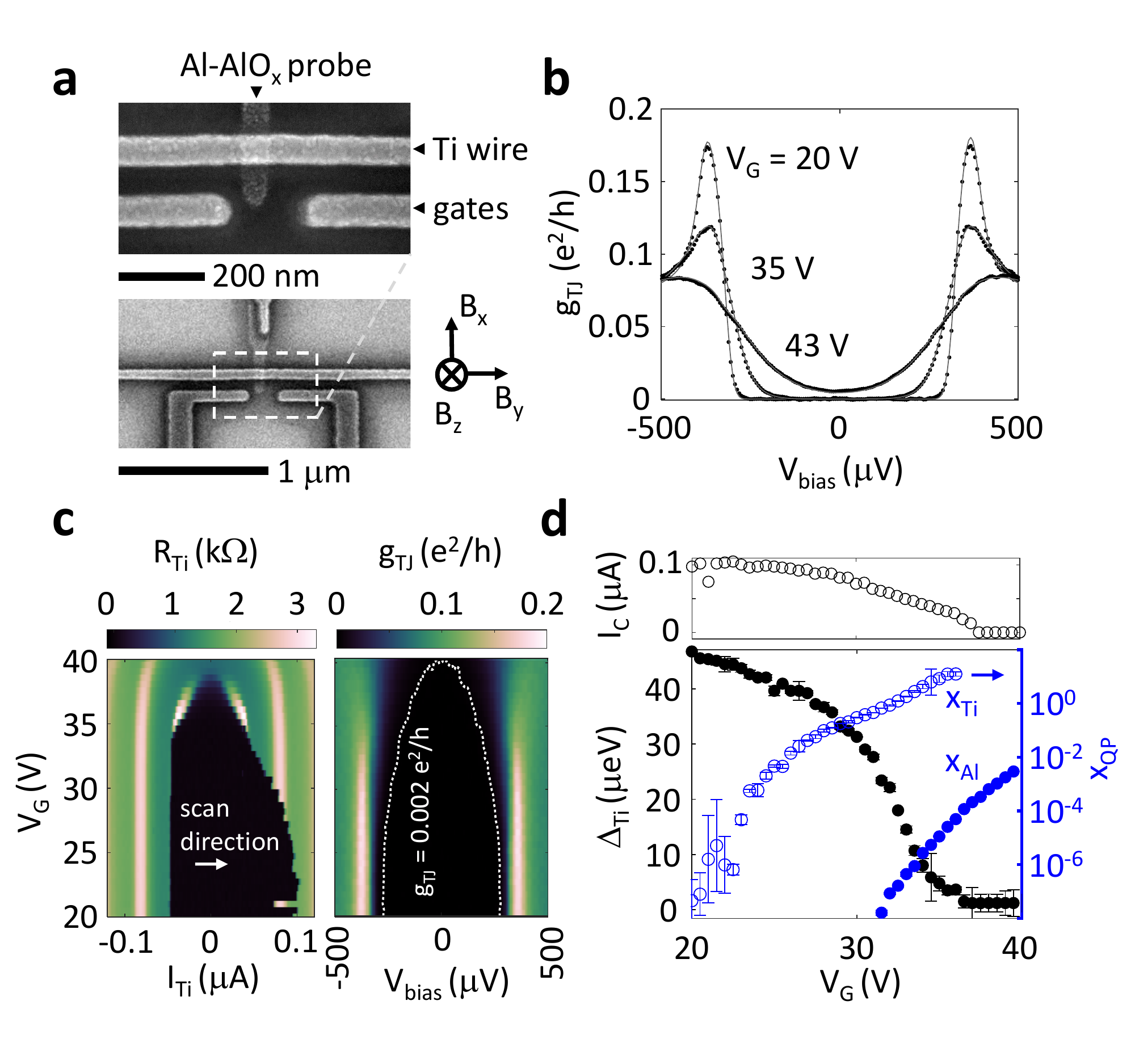}
\caption{\label{fig1} Tunnel spectroscopy and critical current of a titanium nanowire at high gate voltage ($V_G$) and 20 mK temperature. (a) Electron micrograph of the device, which consists of a titanium wire, local surface gates, and an aluminium tunnel probe.  Magnetic field axes relative to device.  (b) Example tunnel data (dots) and fitted Abrikosov-Gorkov model (lines) at the indicated gate voltages. (c) Wire resistance versus current (left) and tunnel spectrum (right) as functions of the applied gate voltage. (d) Gate dependence of the critical current ($I_C$) as compared to the quasiparticle fraction ($x_\text{QP} = n_\text{QP}/n_\text{Cooper pairs}$) and the titanium superconducting gap ($\Delta_\text{Ti}$) inferred from the fitted model. The uncertainties of the fit parameters are calculated as described in the methods section.  The rise in quasiparticle density indicates quasiparticle injection as the mechanism for the observed suppression of the critical current in gated metallic superconductors.
}
\end{figure}

	We fabricate the device shown in Figure 1a via electron-beam lithography and in-situ double-angle shadow evaporation.  We deposit and oxidize a 10 nm thick Al film to form a tunnel probe, followed by a 30 nm Ti film to form a 60 nm wide nanowire channel and nearby gates.  

	Figure 1c shows the critical current and tunnel conductance of the Ti wire as a function of applied gate voltage, measured at 20 mK.  As the bias current is swept, the Ti enters and exits the zero resistance state at the retrapping and critical current ($I_c$) respectively.  Application of gate voltage decreases the critical current before eliminating superconductivity, consistent with the results of \cite{DeSimoni2018}.  In the same device and gate range, we measure the tunnel spectrum (at zero current bias along the nanowire), as shown in the right hand of Figure 1c.  The spectrum broadens with gate voltage, implying an increase of the QP population.

	  We analyze the data using a conventional model for tunnel spectroscopy.  Tunnelling from superconductors can reveal the electronic, magnetic, and phonon structure of materials due to the highly non-linear superconducting density of states\cite{Hauser1971,Worledge2000,Wolf2012}.  The current through a tunnel junction is

 \begin{equation}
\label{ITJ}
I_\text{TJ} = \frac{g_0}{{N_1^\text{(0)} N_2^\text{(0)}}} \int N_1(E) N_2(E+eV_\text{bias})(f(E) - f(E+eV_\text{bias})) dE
\end{equation}
\

where $g_{0}$ is the normal-metal conductance, $N_{1,2}(E)$ are the densities of states for each side of the junction, $N_\text{1,2}^\text{(0)}$ are the normal-metal densities of states, $V_\text{bias}$ is the bias voltage on the tunnel probe, and $f(E) = (1+\exp(E/k_BT_\text{QP}))^{-1}$ is the Fermi function, which defines the QP temperature, $T_\text{QP}$.  The magnetic field serves as a convenient experimental parameter, in the presence of which each density of states is split into spin subbands, $N_{\uparrow\downarrow}$, treated separately in the limit of low spin-flip scattering, and given in the Abrikosov-Gorkov model \cite{Abrikosov1960} as 

\begin{subequations}
\label{Nud}
 \begin{equation}
\label{Nofu}
N_{\uparrow \downarrow} = \frac{N^{(0)}}{2} \text{sgn}(E) \text{Re}(\frac{u_{\pm}}{(u_{\pm} ^2 - 1)^{1/2}})
\end{equation}

 \begin{equation}
\label{us}
u_{\pm} = \frac{E\mp\mu_B B}{\Delta}+ \alpha \frac{u_{\pm}}{(1-u_{\pm} ^2)^{1/2}}
\end{equation}
\end{subequations}
\

where $u_{\pm}$ are defined implicitly, $\mu_B$ is the Bohr magneton, $B$ is the external magnetic field, $\Delta$ is the superconducting gap, and $\alpha$ is the depairing energy.  The depairing energy reflects the typical energy difference between time-reversed electron states.  The functions $N_{\text{(Al,Ti)}(\uparrow,\downarrow)}(E)$ can be obtained analytically by solving equation (\ref{us}) for the $u_{\pm}$.  The total current for both spin channels is then calculated numerically according to equation (\ref{ITJ}), and the derivative with respect to $V_\text{bias}$ gives the conductance 

 \begin{equation}
\label{geq}
g_\text{TJ}(V_\text{bias}) = \left. \frac{dI_\text{TJ}}{dV_\text{bias}}\right|_{\Delta_\text{Al},\Delta_\text{Ti},\alpha_\text{Al},\alpha_\text{Ti},B,T_\text{QP}}
\end{equation}
\

for given gap energies, depairing energies, magnetic field, and quasiparticle temperature.  Moreover, in the approximation that the two layers of the junction are in thermal equilibrium, we can calculate the QP fraction for either layer, $x = \int_0^\infty N(E)f(E)dE/(N^{(0)}\Delta)$. This value is defined as the ratio between the density of occupied excited states, $\int_0^\infty N(E)f(E)dE$, and the density of Cooper pairs, $N^{(0)} \Delta$ \cite{Tinkham1996}.

	The tunnel conductance thus depends on the convolution of the Al and Ti densities of states, causing the peak at the sum-of-gaps seen in Figure 1b-c.  The peak in Figure 1c does not move greatly when $I_C$ becomes zero at $V_G = 37$ V, indicating that the Al gap is much larger than the Ti gap.

	We fit the model of equation (\ref{geq}) to the data of Fig 1c, taking $T_\text{QP}$ and $\Delta_\text{Ti}$ as free parameters, and fixed $\Delta_\text{Al} = 320$ $\mu$eV\cite{Meservey1971}.  The resulting $\Delta_\text{Ti}$ correlates with $I_C$ as plotted in Figure 1d.  We also plot the resulting QP fraction in the Al and Ti layers.  The accuracy of the fit (see Figure 1b) and the adherence of $\Delta_\text{Ti} (T_\text{QP})$ to the self-consistency relation (see Extended Data Figure 1) indicate that QP injection, rather than a change in the underlying electronic or crystal structure, destabilizes superconductivity.

	Still, it might be argued that a magnetic mechanism may lead to broadening like that observed.  For instance, under high electric fields, oxygen ions might accumulate and present a large moment at the surface of the Ti \cite{Bi2014, Kumar2016}.

\begin{figure}[h!]
\includegraphics[width=4 in]{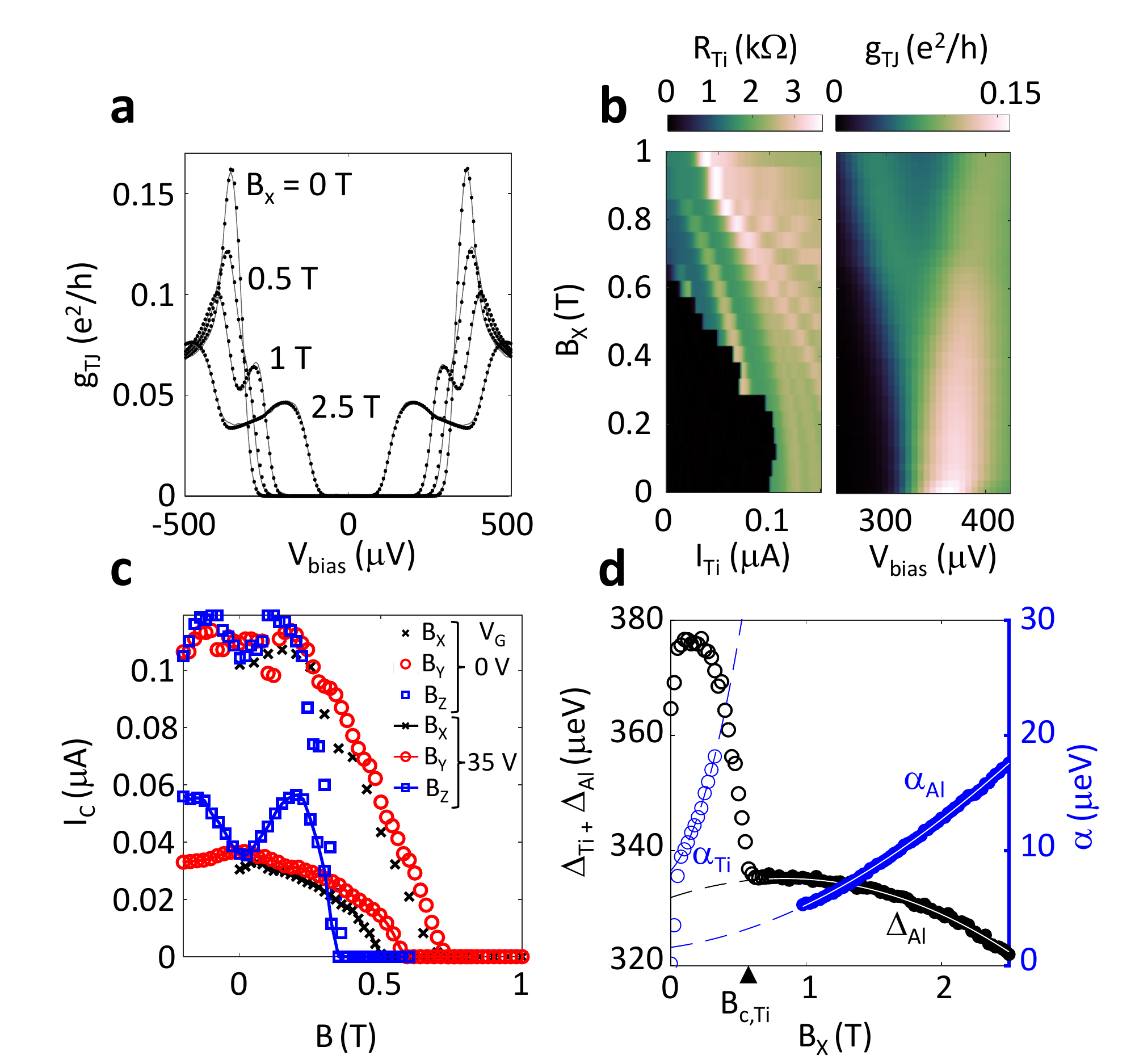}
\caption{\label{fig2} Device characterization in applied magnetic field.  (a) Example tunnel data (dots) and fits (lines) with magnetic field applied along the x-axis, illustrating the Zeeman splitting of the density of states and the depairing effect, which is distinct from the effect of quasiparticle injection. (b) Comparison of the titanium resistance (left) and the tunnel spectrum (right) as functions of the magnetic field. (c) Titanium critical current under $x$, $y$, and $z$ field orientations, both at zero and elevated gate voltage. (d) The gap and depairing energies ($\Delta_\text{Al,Ti}$, $\alpha_\text{Al,Ti}$,) obtained by fitting the tunnel data above (filled circles) and below (open circles) the titanium critical field ($B_\text{c,Ti}$) as described in the text.
}
\end{figure}

	Figure 2 explores the effects of magnetic field to characterize the films and consider the possibility of a magnetic gating effect.  Figure 2a shows the spectrum at several values of magnetic field.  The model accurately captures the orbital depairing and Zeeman splitting.  The critical current data (Figure 2b, left) show a critical magnetic field in the Ti at $B _x = 0.7$ T, and inspection reveals a discontinuity at this field in the respective tunnel data (Figure 2b, right).  To fit the model to the tunnelling data, we first consider the region 1 T $<$ $B _x$ $<$ 2.5 T, in which the Ti is normal, and the Al superconducting.  Here, the model has free parameters $\Delta_\text{Al}$ and $\alpha_\text{Al}$ which follow quadratic trajectories shown as dashed lines in Figure 2c ($T_\text{QP} = 20$ mK).  These trajectories are used to fit the low-field region, where only $\Delta_\text{Ti}$ and $\alpha_\text{Ti}$ are taken to be free parameters.  The resulting Ti parameters also follow quadratic dependences, and from the relation $\alpha = (e H_{\parallel}^2 d^2/6\hbar + H_{\perp}) D e / c$ for a thin film of thickness $d$, we find the diffusion constants $D_\text{Al} = 0.6$ cm$^2$s$^{-1}$ and $D_\text{Ti} = 2.4$ cm$^2$s$^{-1}$ with a 2.5 degree field misalignment\cite{Tinkham1996}.  The diffusion constants are typical of thin polycrystalline metal films\cite{Meservey1971}.  Further, Figure 2c shows the critical current for magnetic fields along all device axes and for both zero and high gate voltage, illustrating the predominantly isotropic nature of the gate effect (complete data in Extended Data Figure 2).  If we try to account for the broadening due to the gate (Figure 1b) with the depairing parameter $\alpha$ rather than $T_\text{QP}$, the best fit does not match the lineshape of the data (see Extended Data Figure 3).  By distinguishing between the effect of magnetic fields ($\alpha$) and the effect of the gate voltage ($T_\text{QP}$) we exclude any pair-breaking origin of the gate effect that is ergodic in the sense defined by de Gennes \cite{Tinkham1996}.

\begin{figure}[h!]
\includegraphics[width=4 in]{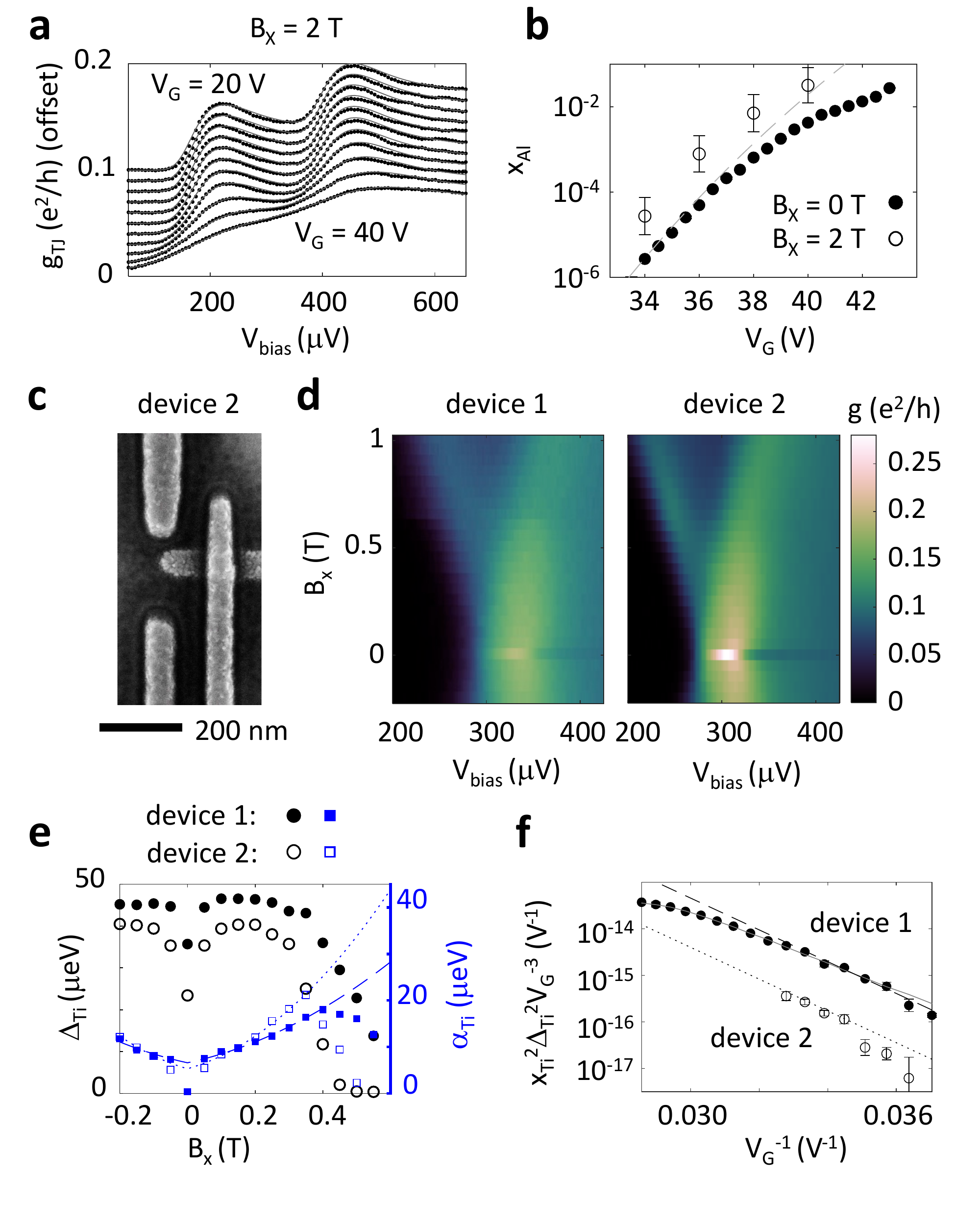}
\caption{\label{fig3} Injection in fixed high magnetic field, a second device, and a model of quasiparticle injection due to field emission.  (a) Tunnelling spectra $g_\text{TJ}$ vs $V_\text{bias}$ for  gate voltages $V_G$ from 20 to 40 V (2 V intervals) with fixed $B _x = 2$ T, well above the titanium critical field. (b) The calculated quasiparticle fraction $x$ within the aluminium for $B _x = 2$ T (open circles), $B _x = 0$ (filled circles), and the function $a\exp (-V_0/V_\text{G})$ (dashed line). The error bars are the uncertainties calculated in the fitting procedure as detailed in the methods section. (c) A second, `tunnelling-only' device displaying a sharper spectrum (d) but similar magnetic field (e) and gate voltage dependences to the first device. (f) The observed quasiparticle fraction is largely explained by a simple model of quasiparticle dynamics assuming Fowler-Nordheim emission from the gate (dashed and dotted lines).  A more detailed model (grey curve) incorporates one-dimensional diffusion away from the injection region.
}
\end{figure}

	Next, in Figure 3a, we apply a fixed field $B_x = 2$ T such that the Ti is in the normal state, and then apply gate voltage.  These data admit simpler analysis: with the Ti normal, equation (\ref{geq}) reduces to $g_\text{TJ}(V_\text{bias}) \sim \int N_{Al}(E-eV_\text{bias})f'(E-eV_\text{bias})dE$.  The spectra directly represent the thermally-broadened Al density of states.  The observed broadening with $V_G$ confirms the hypothesis of QP generation, and is largely independent of applied magnetic field.

	We note two differences between this experiment and classical QP injection experiments\cite{Levine1968,Smith1975,Ullom2000}.  First, the injected QP originate at energies ($\approx eV_G$) inaccessible to the spectroscopy, and, second, the injection current is so low in our case it can be difficult to measure.  (The model below indicates a current of at most 0.1 nA, see Extended Data Figure 4.)  Nonetheless, we can examine $x(V_G)$ for clues to the underlying transport mechanism.  We plot $\log(x_\text{Al}(V_G))$ in Figure 3b, as calculated from the data of Figure 3a and 1c.  The Al data roughly follow $x\sim \exp(-V_0/V_G)$ for $V_0$ = 2 kV, which suggests field emission.  But, in our geometry, current from the gate most directly enters the Ti.  Consequently, we focus on the QP population in the Ti within the following simple kinetic model.  


Given an injection current $I_G$ the number of QP in the Ti portion of the device is approximately

\begin{equation}
\label{nexpr}
n_{\text{QP}} = \frac{I_G V_G \tau_\text{eff}}{\Delta}
\end{equation}
\

where $\tau_\text{eff} = \tau_0/x$ is the effective QP lifetime\cite{Levine1968,Barends2008}. That is, each electron from the gate excites $V_G/\Delta$ thermal quasiparticles which decay in proportion to their density due to recombination.  In terms of $x$, the value $n_{\text{QP}}= x \Omega N^{(0)}\Delta_\text{Ti}$, where $\Omega$ is the volume of the injection region.  The current between two metals separated by a vacuum can be calculated from the Fowler-Nordheim equation, $I = a V^2 \exp(-b/V)$.  Here $a = e S / (16 \pi^2 \hbar \phi l^2)$ and $b = 4 l (2m\phi^3)^{1/2} / (3\hbar$) with $S$ the emission area, $\phi$ the work function, and $l$ the separation\cite{Jensen2018}.  Substituting into equation (\ref{nexpr}), we expect the following relation: 

\begin{equation}
\label{nexpr2}
\frac{x^2 \Delta^2}{V_G^3} = A e^{-b/V_G} 
\end{equation}

\

with $A = a \tau_0 / (N^{(0)} \Omega)$.  In Figure 3f, we plot the left-hand quantity (from the data of Figure 1c) on a log scale versus $V_G^{-1}$ and find a nearly linear dependence.  The dashed line corresponds to $l = 33$ nm and $\phi$ = 2.3 eV.  The value of $A = 6\times10^{-4}$ V$^{-1}$ accords with the realistic estimates $\tau_0 = 5$ ns, $S = 30$ nm $\times$ 0.6 $\mu$m, $N^{(0)} = 7\times10^{10}$ $\mu$m$^{-3}$eV$^{-1}$, $\Omega = 4$ $\mu$m $\times$ 60 nm $\times$ 30 nm, validating the field emission hypothesis and indicating that the current impinges on the superconductor with the energy of the electrode.  A model taking into account QP diffusion along the wire is described in Extended Data Figure 5, and captures the nonlinearity in Figure 3f.  

	To confirm the results, we study a second device in Figure 3c, which is similar to the first except that the Ti wire is terminated on one side.  In Figure 3d we compare $g(V_\text{bias}, B)$ of the two devices and attribute the comparative sharpness of the second device to microscopic differences in the films.  The results of fitting the tunnel data in applied magnetic field and electric field are similar to those of the first device (Figure 3e-f).  The higher diffusion constant, $D_\text{Ti} = 3.2$ cm$^2$s$^{-1}$, in the second device may contribute to its having a lower $A$ parameter ($A = 6\times10^{-5}$ V$^{-1}$).   In addition, several devices with simpler geometries were measured to directly reproduce the results of \cite{DeSimoni2018} (see Extended Data Figure 6).  In all devices, the effect is found to be stable over week-long time-scales.

	With sufficient electric field, mesoscopic superconductors can be switched into the normal state with very low quiescent currents.  Introducing a tunnel junction, we have directly observed the governing process in which field emission generates quasiparticles.  The effect may be suited as a generally applicable switch of superconductivity, since field emission minimizes dissipation outside of the channel, in contrast to prior three-terminal superconducting devices\cite{McCaughan2014}.  For quantum information processors, voltage-controlled coupling resonators would have more scalable interconnects so long as the QP introduced do not limit the overall circuit's coherence.  Finally, the technique extends the experimentally accessible realm of superconductor relaxation processes to higher energy and nanometer scale perturbations, meriting study in other materials.  For example, the QP source could be used to evaluate new designs of local QP traps or phonon barriers, as have been pursued recently in the contexts of quantum information and photon detectors.

\section{Methods}

The devices are fabricated beginning with a bilayer resist (950 kDa PMMA, 4\% in anisole, on (MMA (8.5) MAA) copolymer, 11\% in ethyl lactate) on dry-oxidized (90 nm SiO$_2$), degenerately doped (001) Si, which is exposed using a 150 kV electron beam pattern generator to create a suspended mask.  10 nm of Al is thermally evaporated at 30 degrees from normal incidence and oxidized in 100 Torr medical grade air for 30 min, after which 30 nm Ti is deposited at normal incidence.  The devices are cleaned by UV-ozone treatment (Samco UV-1, 40 $\deg$ C, 3 min) after acetone liftoff. The zero-field values of the gaps, $\Delta_{\text{Al},0} = 320$ $\mu$eV and $\Delta_{\text{Ti},0} = 50$ $\mu$eV, are consistent with typical enhancement of superconductivity in environmentally exposed Al films and degradation of Ti relative to bulk materials ($\Delta_{\text{Al,bulk}} = 180$ $\mu$eV and $\Delta_{\text{Ti,bulk}} = 75$ $\mu$eV)\cite{Meservey1971}.  The diffusion constants measured in the main text correspond to mean free paths $l \approx 3 D/v_\text{F}$ of 1 - 5 \AA, \ consistent with near-unity residual resistance ratios in these films (RRR$_\text{Ti} = 1.3$, RRR$_\text{Al} = 1.05$) \cite{Hsieh1968,Meservey1971}.  Four terminal measurements of the devices are performed in a dilution refrigerator at 17 - 22 mK equipped with low-temperature low-pass filtering, using a Keithley 2400 as gate voltage source. In order to tolerate a slight misalignment in the shadow evaporation process, the gate voltage is applied to the right hand gate in figure 1a, as detailed in Extended Data Figure 4.

To model a given sequence of tunnelling data ($g_\text{TJ}$ versus either $B$ or $V_G$), 400 random initial points in the parameter space are taken, from each of which a gradient descent is performed.  The objective function is calculated with 0.1 $\mu$eV resolution from equation (\ref{geq}).  Subsequent fits start from 100 initial parameter points near to the best of the previous optimization.  The results are found to be robust to constraints and details of the fitting procedure.  The uncertainties in the fit parameters are typically 2 $\mu$eV and are calculated by the delete-$m$ jackknife method.   The spectroscopy below the critical field of the Ti consistently contains a fixed normal metal component which we attribute to inhomogeneities in the Ti and incorporate into the model as detailed in Supplementary Figure 1.  Even at the lowest temperatures, the line-width of $N_\text{Al}$ exceeds $k_B T$ to an extent attributable to gap anisotropy in Al\cite{Wolf2012}. This is accounted for in the model by uniformly convolving the $N_\text{Al}$ with a normalized Gaussian of width 18 $\mu$eV for device 1 and 9 $\mu$eV for device 2.  The simple convolution does not introduce significant error as compared to a more realistic model in which the gap is distributed uniformly according to the known gap anistropy in Al as shown in Supplementary Figure 2\cite{Blackford1976}.

\section{Acknowledgements}
We thank Pat Gumann and Anthony Annunziata for discussions. This work is supported by IBM Quantum, under Q Network for Academics program. This work is also partly supported by NSF grant DMR-1708688.  LDA acknowledges support from an appointment to the Intelligence Community Postdoctoral Research Fellowship Program at Harvard University, administered by Oak Ridge Institute for Science and Education through an interagency agreement between the U.S. Department of Energy and the Office of the Director of National Intelligence.

\section{Author Contributions}
LDA, AKS, CGB, ATP, SHL, and SPH performed low-temperature measurements:  LDA and AKS fabricated the devices and performed the data analysis: LDA and AY designed the experiments:  All authors discussed the results and commented on the manuscript.

\section{Data Availability}

The transport data and device images that support the findings of this study are publicly available in the Harvard Dataverse with the identifier doi:10.7910/DVN/LHCDHV 

\section{Additional Information}

Supplementary information is available in the online version of the paper. Reprints and permission information is available online at www.nature.com/reprints. Correspondence and requests for materials should be addressed to LDA.

\section{Competing Interests}

The authors declare no competing interests.

\renewcommand{\figurename}{Extended Data Figure}
\renewcommand{\thefigure}{\arabic{figure}}
\setcounter{figure}{0}

\begin{figure}[h!]
\includegraphics[width=3.5 in]{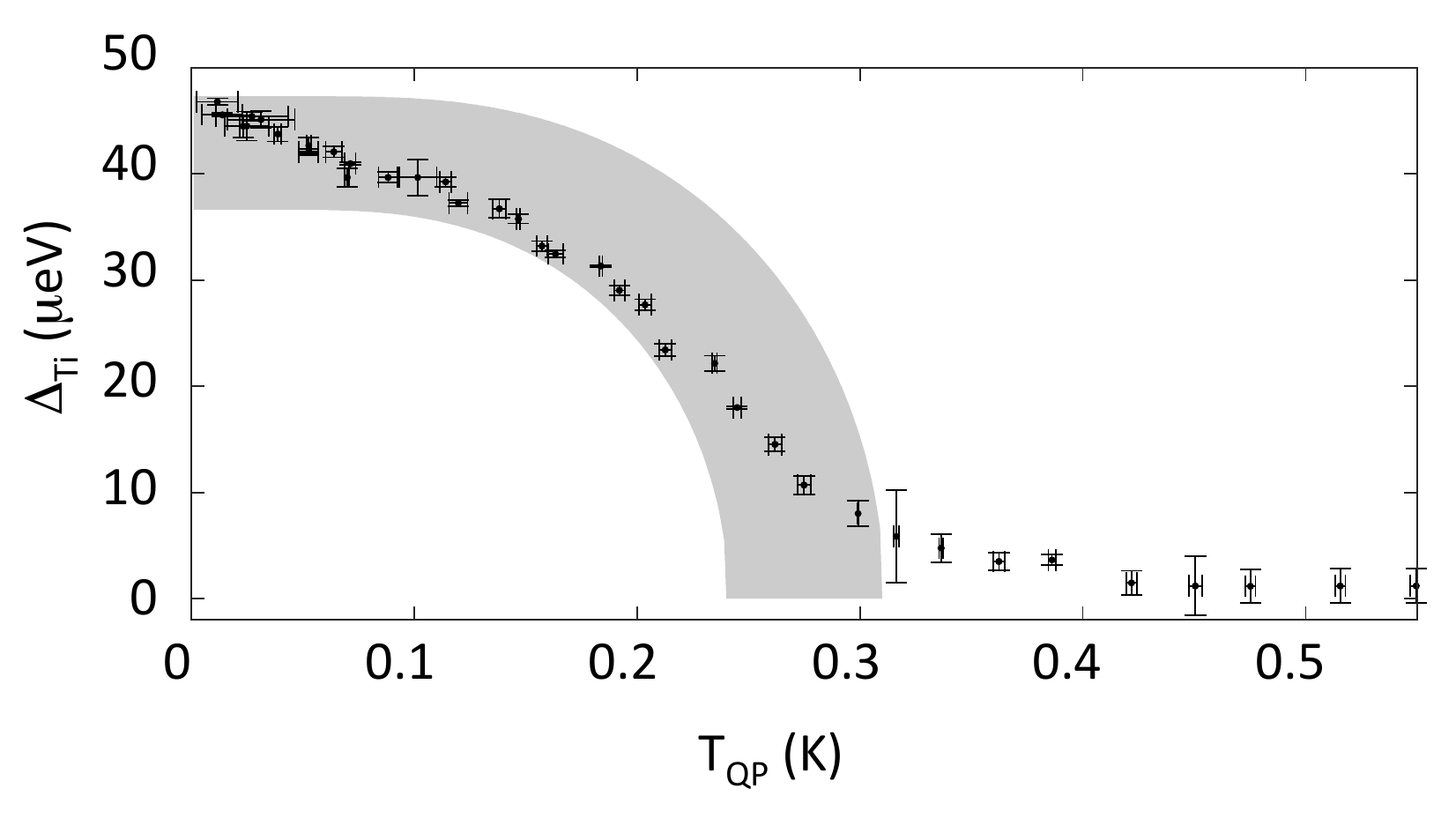}
\caption{\label{ext1}
Comparison of the spectroscopy fitting to the self-consistency relation for weak-coupling superconductors. This relation links the zero temperature gap to the critical temperature as $\Delta_0= 1.764$ $k_\text{B} T_c$, and the temperature dependence of the gap implicitly through the relation $\text{ln}(2e^\gamma \hbar \omega_D /k_\text{B} T_c \pi) = \int_0^{\hbar \omega_D} \text{tanh}(\sqrt{\xi^2 + \Delta(T)^2}/2 k_\text{B} T)/\sqrt{\xi^2 + \Delta(T)^2} d\xi$ where $\gamma$ is Euler's constant, and $\hbar \omega_D$ is the Debye energy.  For weak coupling $\hbar \omega_D \gg k_\text{B} T$, so that $\Delta(T)$ is parametrized only by $T_c$ [23].  Above we plot the titanium gap energy versus $T_{QP}$ extracted from the spectroscopy model (i.e. from the data of Fig. 1c-d).  The data are within the range corresponding to $T_c =$ 0.24 - 0.31 K, the grey region, obtained by solving the foregoing equation numerically for those two bounds.  The conformity to the quasiparticle population to the self-consistency relation is therefore quite good, which further excludes an exotic dissipationless gate effect.  The slight departure of the data from a typical BCS dependence may be due to variations in the non-equilibrium state due to the energy of the impinging electrons, which varies dramatically over this data set. The error bars are calculated as described in the methods section.
}
\end{figure}

\begin{figure}[h!]
\includegraphics[width=6 in]{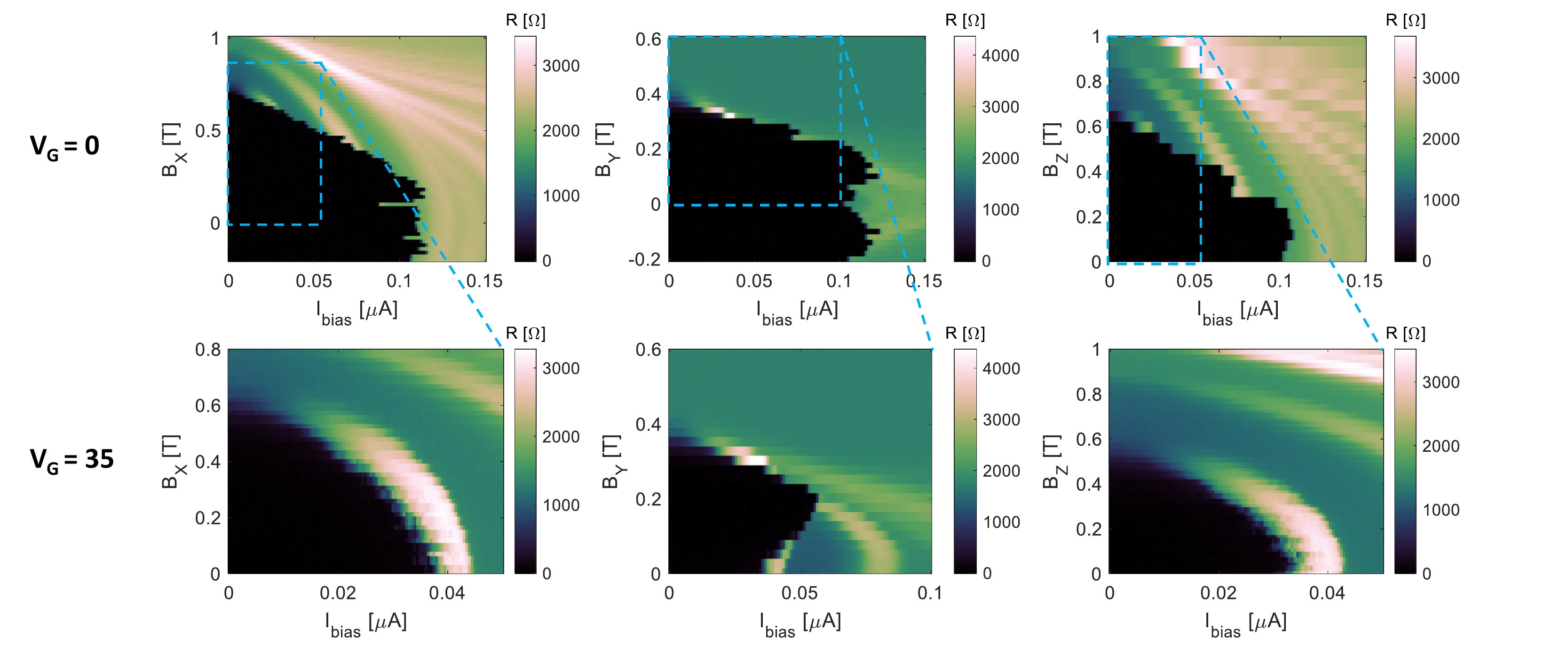}
\caption{\label{ext2} Titanium wire magnetotransport.  Resistance of the Ti wire of Figure 1 as a function of magnetic field applied along $x$, $y$, and $z$ directions for zero applied gate voltage (top row) and  35 V (bottom row).
}
\end{figure}

\begin{figure}[h!]
\includegraphics[width=3.5 in]{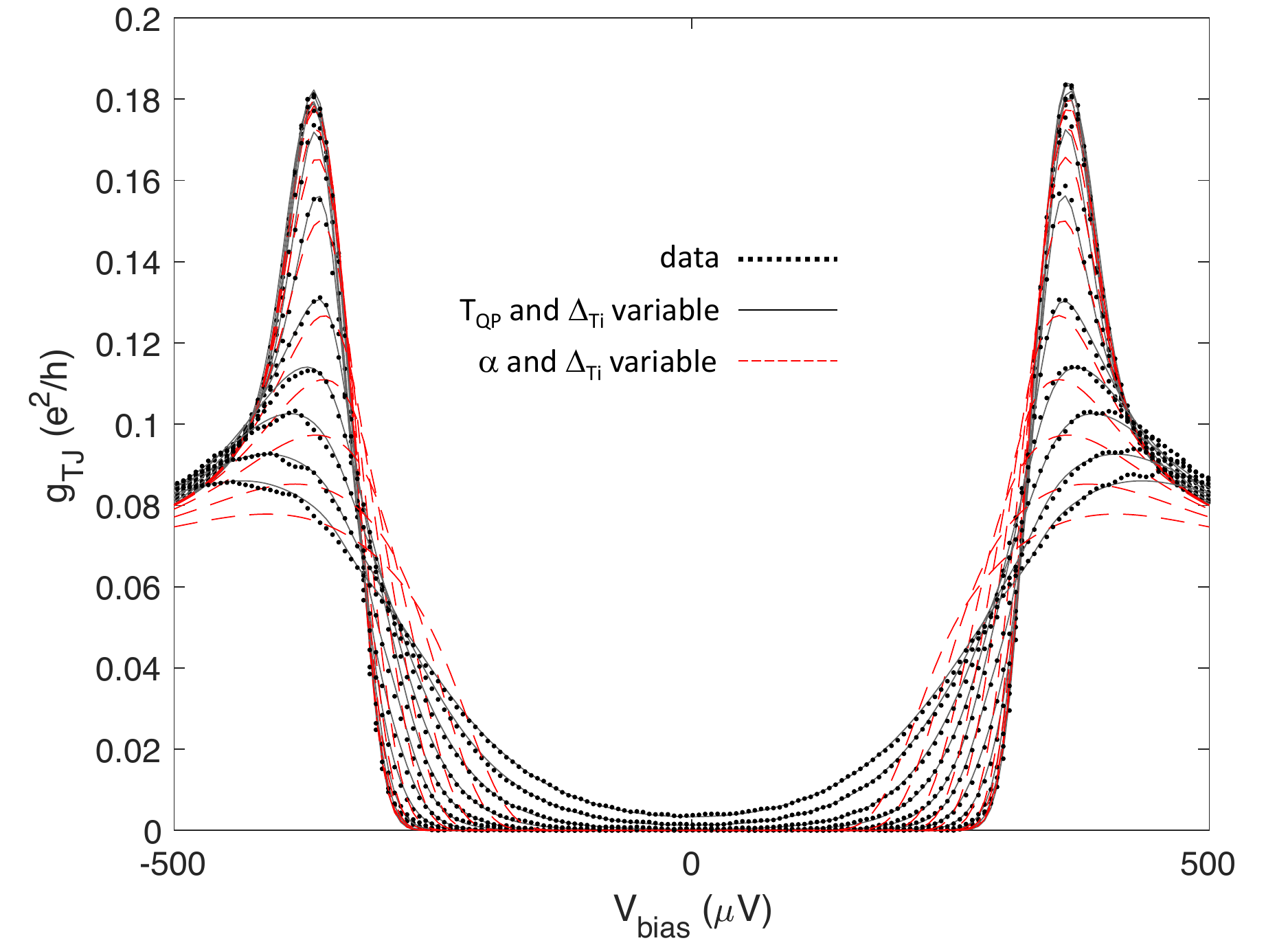}
\caption{\label{ext3} Broadening mechanism comparison.  Tunnel junction conductance vs bias data (dots) for gate voltages from 20 V (most peaked) to 43 V (least peaked) as compared to best fits with $T_\text{QP}$ as free parameter (lines) and best fits with $\alpha_\text{Al}$ as free parameter (dashed lines).
}
\end{figure}

\begin{figure}[h!]
\includegraphics[width=6 in]{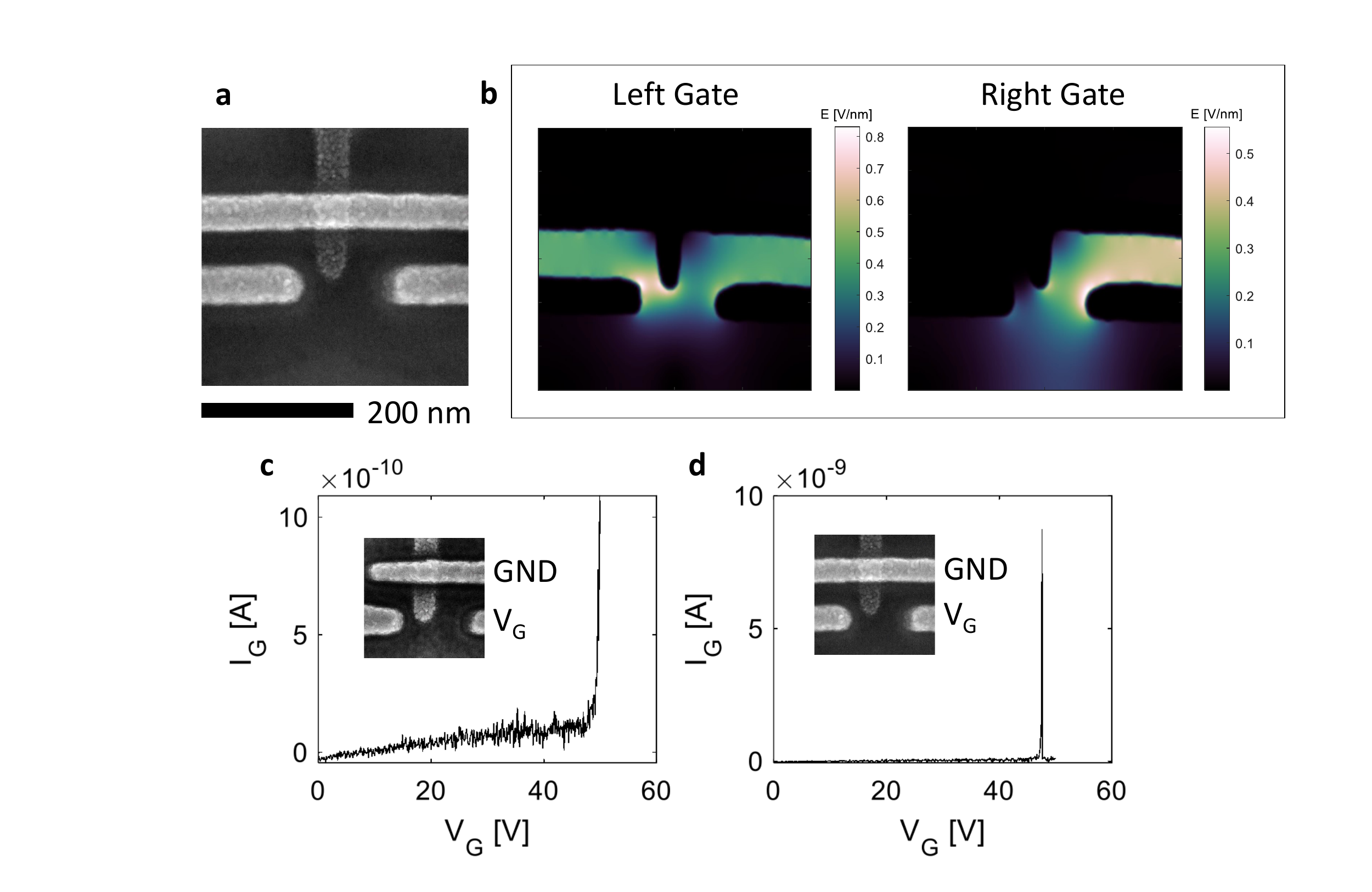}
\caption{\label{supp1}  Electric field calculation and current measurements.  The device of the main text (a) has a slight asymmetry in the $y$-direction.   As a result, the left and right gates produce different electric field distributions, the magnitude of which we calculate numerically in (b) for 40 V applied to either gate.  As a result of this asymmetry, we use the right gate in the data of the main text, since this most effectively applies field to the Ti.  (c) To look for current flow through the gate, we measure gate current in devices identical to the devices in the text, but at 4.2 K.  We observe breakdown at 45 - 60 V in such devices, above the region of stable emission in the text.  Below breakdown, the small amount of current detected is ohmic and almost certainly takes place in the contacts in this measurement.  In the devices of Extended Data Figure 6, measured in a separate dilution refrigerator with high line-isolation, current was measured to be less than 1 pA for $V_G$ = 52 V, but these devices had a different gate geometry (90 nm SiO$_2$ back gate).  The Fowler-Nordheim model described in the main text predicts current of $\sim$ 0.1 nA at the highest voltage.  
}
\end{figure}

\begin{figure}[h!]
\includegraphics[width=6 in]{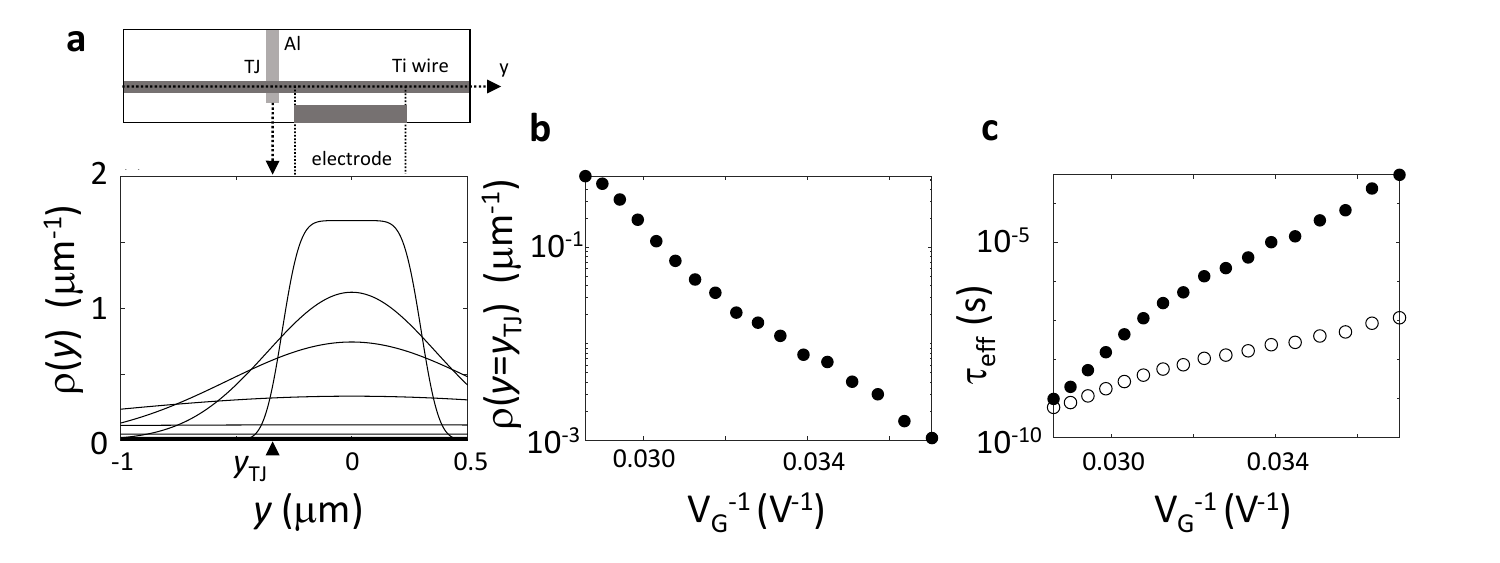}
\caption{\label{ext4}
1D quasiparticle diffusion model. The minimal model of quasiparticle dynamics presented in the main text does not account for diffusion of QP along the Ti wire, away from the injection region.  Correcting this can account for the departure of the data from equation (\ref{nexpr2}) as follows:  We assume that the QP distribute along the length of the Ti wire according to $\rho(y)$ due to one-dimensional diffusion over a distance $(2 D_\text{Ti} \tau_\text{eff})^{1/2}$ where $D_\text{Ti}$ is the diffusion constant calculated from the magnetic field data.  That is $\rho(y) = \rho_0 \int_{-l_I/2}^{l_I/2} \exp((s-y)^2/2D_\text{Ti}\tau_\text{eff}) ds$ where $\rho_0$ is such that $\int \rho(y) dy = 1$, and $l_I$ is the injector length. Moreover, we relax the assumption that $\tau_\text{eff} \sim x_\text{Ti}^{-1}$. The new model amounts to replacing $A$ on the right hand side of equation (\ref{nexpr2}) with $A \rho(y_\text{TJ},\tau_\text{eff}) x_\text{Ti} \tau_\text{eff} \Omega/(l_I t_\text{TJ} w_\text{TJ} )$ where $y_\text{TJ}$ is the distance from the tunnel junction to the centre of the injection electrode, $l_I$ is the length of the injection electrode, $t_\text{TJ}$ is the thickness of the tunnel junction portion of the Ti wire, $w_\text{TJ}$ is the width of that portion, and now $\tau_\text{eff} = \tau_0 x^\nu$.  The values $l_I$ = 600 nm, $t_\text{TJ}$ = 5 nm, $w_\text{TJ}$ = 7 nm, and $\nu = 2.5$ correspond to the curved line in Figure 3f of the main text, which closely follows the data.  However, the junction dimensions here are lower than expected, and a finite element model would be still more realistic.  (a) $\rho(y)$ plotted at the relevant gate voltages, with the sharpest distribution corresponding to the highest voltage. (b) $\rho(y)$ evaluated at the location of the detector as a function of the gate voltages of Figure 3f. (c) The effective lifetime of quasiparticles implied by this model (filled circles) as compared to the minimal model described in the main text (open circles).  
}
\end{figure}

\begin{figure}[h!]
\includegraphics[width=6 in]{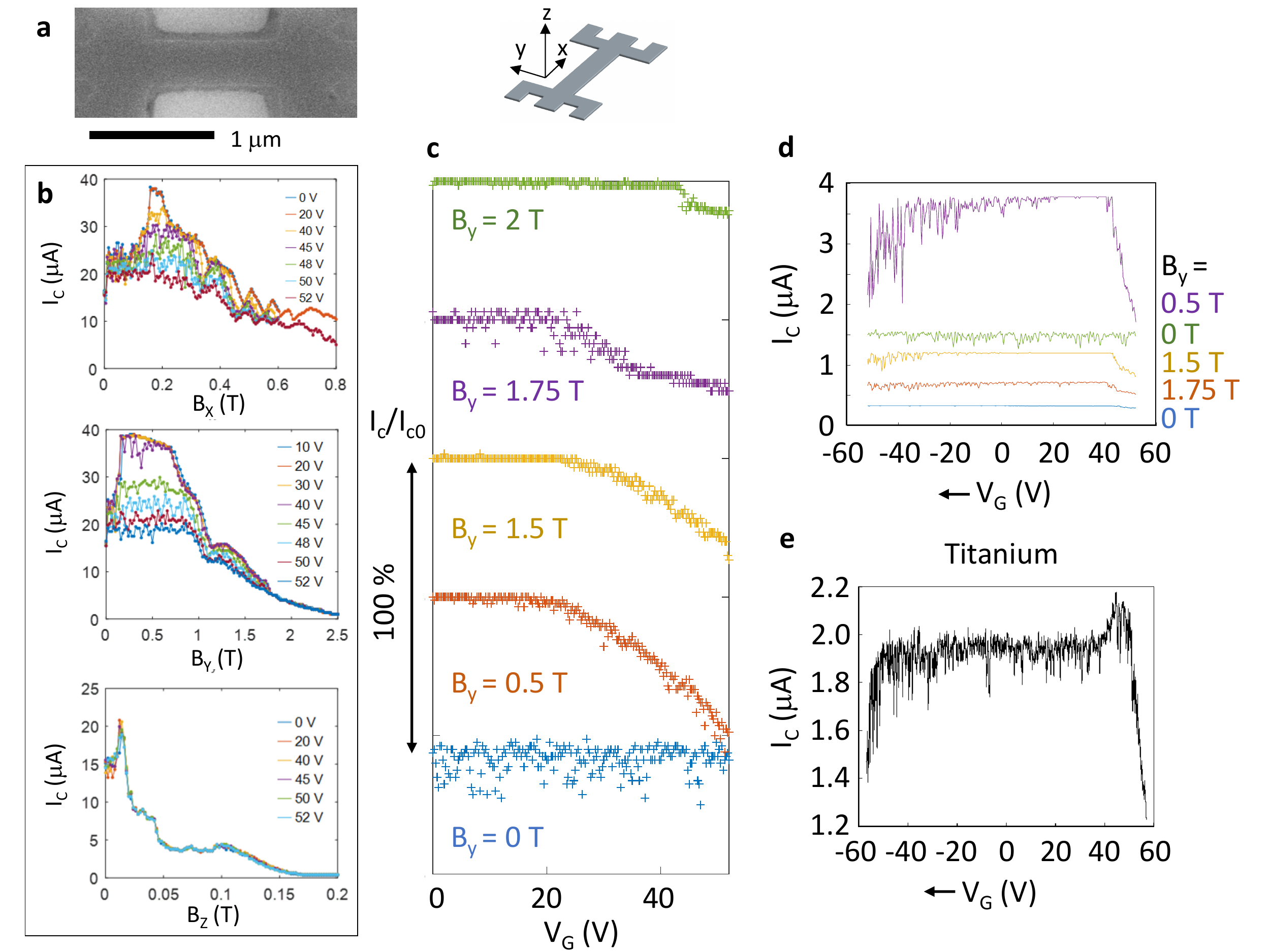}
\caption{\label{ext5}  Additional Al and Ti devices.  Wide, back-gated critical current devices are measured at 50 mK in the geometry shown in (a).  In a device composed of a 10 nm Al film on a 90 nm SiO$_2$ gate dielectric, the gate effect is observed only at elevated magnetic field.  Critical current measurements are performed both at fixed gate voltage while sweeping magnetic field (b), or at fixed magnetic field while sweeping gate voltage (c). Oscillations observed at high $B_x$ are likely to be related to Weber blockade and can be adjusted by the gate voltage. Hysteresis is observed in the gate effect, as can be seen in (d) in which the gate is swept from high to low voltage (over 20 minutes).  A reversed behavior occurs when the gate is swept from low to high.  (e) A similar Ti device shows a gate effect at zero magnetic field and the hysteresis takes the form of an overshoot of the gate effect (peak at positive $V_G$).  The thickness of the Ti here is 30 nm.
}
\end{figure}

\renewcommand{\figurename}{Fig.}
\renewcommand{\thefigure}{S\arabic{figure}}
\setcounter{figure}{0}

\end{document}